# Paper 04

# An Event-driven Operator Model for Dynamic Simulation of Construction Machinery

Reno Filla

VOLVO WHEEL LOADERS AB, ESKILSTUNA, SWEDEN

**Abstract**

Prediction and optimisation of a wheel loader's dynamic behaviour is a challenge due to tightly coupled, non-linear subsystems of different technical domains. Furthermore, a simulation regarding performance, efficiency, and operability cannot be limited to the machine itself, but has to include operator, environment, and work task.

This paper presents some results of our approach to an event-driven simulation model of a human operator. Describing the task and the operator model independently of the machine's technical parameters, gives the possibility to change whole sub-system characteristics without compromising the relevance and validity of the simulation.

*Every passing minute is another chance to turn it
all around.*

*(from the film "Vanilla Sky")*

# 1 Introduction

In the development of off-road machinery, dynamic simulation of large and complex technical systems is being increasingly practised. In the case of a wheel loader, most subsystems are non-linear and tightly coupled, which makes prediction and optimisation of the complete system's dynamic behaviour a challenge. Both drive train and hydraulics are competing for the limited engine torque. As described in [1], the momentary power distribution is specific for the task at hand and is controlled by the operator, who ultimately balances the complete system. Figure 1 shows how power is transferred through all relevant wheel loader subsystems, with the machine being used in the typical work task of loading gravel.

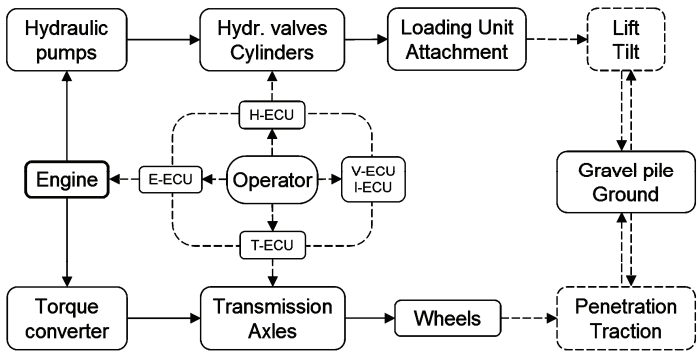

**Figure 1.** Simplified power transfer scheme of a wheel loader loading gravel

So-called full vehicle models can be useful in various types of simulation. How the model needs to be controlled and subjected to virtual loads is very much dependent on the purpose of the simulation. Paper [2] gives examples and reasons of different approaches.

Our work aims to simulate a wheel loader's total performance, efficiency, and operability, and to investigate the robustness of these three important product properties. Because a human operator adapts to the machine and the work situation, assessing the above mentioned properties by means of simulation also requires modelling operator, environment, and task at an appropriate level of detail.

Traditional backwards simulation forces a technical system to follow a prescribed cycle, which can result in physically erroneous results. Forward simulation with fixed operator input would use each machine in the same way, producing physically correct but irrelevant results. An experienced operator can adapt to a new machine in order to achieve the highest possible performance. Being able to simulate this adaptation to a certain degree is a necessary first step towards optimisation of the complete system.

Paper [2] describes the initial results of our work on an operator model, with a focus on the bucket filling phase. Since the complete loader was modelled in a program for multi-body simulation (Figure 2), our first approach was to also realise the operator model in this simulation package. As noted in the discussion section of that paper, this proved to be possible but cumbersome. It seemed that for this type of problem, realisation as a finite state machine in a discrete-event simulation program (in co-simulation with the multi-body simulation package) would have been better.

Consequently, this was our next step. Choosing a program suite that supports both continuous and discrete event simulation, the operator can now be modelled with a variety of control strategies, completely separate from the machine. A simulation can be performed as either co-simulation (where both sides solve their own equations and communicate the results) or as a plant import into the multi-body simulation program (where the operator model has been compiled into an executable file, which is referenced as a General State Equation).

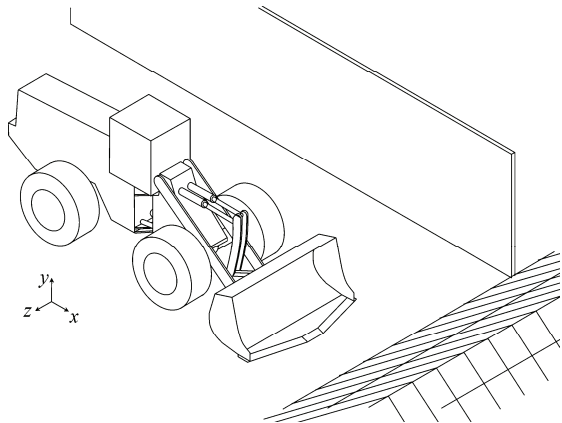

**Figure 2.** Models of the wheel loader and its environment

With the bucket filling phase already covered in [2], this paper focuses on the remaining essential elements of a wheel loader working cycle.

## 2  Loading Cycle Description

Wheel loaders are highly versatile machines, and with each task and workplace being unique, it is difficult to define a standard test cycle that covers every possible aspect. However, tasks do exist that are more common than others. Figure 3 shows a so-called *short loading cycle*, sometimes also dubbed *V-cycle* or *Y-cycle* for its characteristic driving pattern. Typical for this cycle is the loading of some kind of granular material on an adjacent load receiver (e.g. a dump truck).

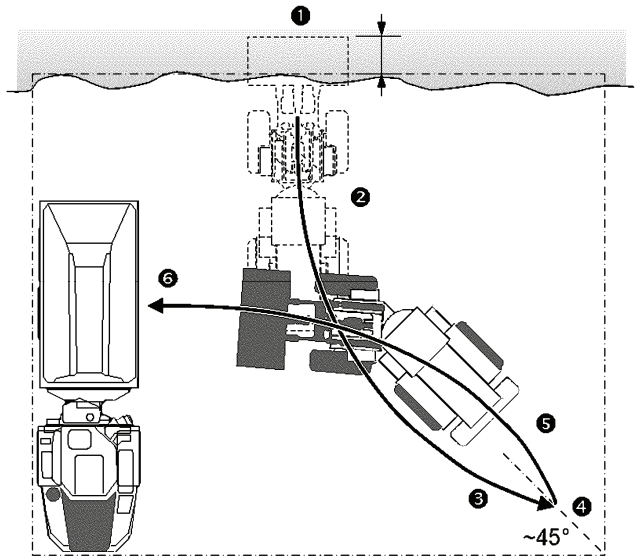

**Figure 3.** Short loading cycle

In paper [3], such a short loading cycle is divided into different phases, and some examples are given for especially interesting situations that occur in some phases.

Phase 1, *Bucket filling*, begins as soon as the bucket's cutting edge is in contact with the material. The operator then uses the kick-down function to shift into the lowest gear. Lifting the loading unit somewhat increases force on the front axle, which improves traction. The operator then simultaneously controls the machine speed (via the engine throttle) and lift and tilt function (via hydraulic levers) in order to fill the bucket. Again, this has already been covered in [2] and will not be considered in this paper.

Phase 2, *Leaving bank*, begins after the bucket is fully tilted back. The lift function is activated, the transmission is put into reverse, and the engine speed is increased. After leaving the pile, the machine is steered to the side in the characteristic V-pattern. The lifting function is activated all the time until reaching the load receiver at the end of phase 5.

Phase 3, *Retardation*, begins when the operator judges that the remaining distance to the load receiver is sufficient for the lift hydraulics to accomplish the necessary bucket height while driving (operators usually prefer driving a longer distance to waiting in front of the load receiver for a sufficient bucket clearance). Most operators use engine braking for retardation.

*Reversing* in phase 4 is unfortunately usually accomplished by putting the transmission into forward while the machine is still moving backwards. This is convenient, but increases fuel consumption, since the torque converter's turbine wheel is forced to rotate backwards until enough torque has been built up and the machine is moving forwards.

Phase 5, *Towards load receiver*, begins when the machine speed is positive, i.e. the machine is moving forward. The operator steers the loader to achieve the V-pattern that is characteristic for a short loading cycle. At the end of this phase, the loader arrives perpendicularly at the load receiver. If the operator judged correctly in phase 3, then the bucket has reached a sufficient height to begin unloading. Usually this is the case, because an experienced operator is able to adapt to any machine very quickly.

In phase 6, *Bucket emptying*, the operator often drives forward very slowly while at the same time raising the loading unit and tilting the bucket forward. This gives the possibility to not just dump the material on the load receiver, but actually place it so that after 3 to 4 buckets a dump truck is evenly loaded without any material being spilled.

In paper [3], the short loading cycle continues with phases 7-10: *Leaving load receiver*, *Retardation and reversing*, *Towards bank*, and *Retardation at bank* (the last is often combined with the next bucket filling by using the machine's momentum to drive the bucket into the gravel pile, rather than applying traction with the wheels). In this stage of our work, however, only phases 1-6 are covered, because of the significant interaction between all main subsystems.

The challenge in such a short loading cycle, from a manufacturer's point of view, is to find an appropriate, robust, and maintainable balance between productivity (material loaded per time), efficiency (fuel consumed per load), and operability over the complete cycle.

Productivity and efficiency are well-defined, but a generally agreed definition of operability has still to be found. In [4], a definition is offered, that also works well even for construction machinery: "Operability is the ease with which a system operator can perform the assigned mission with a system when that system is functioning as designed". The limitation to states where the system is functioning as designed, effectively excludes properties like robustness and reliability. But this still gives no explanation as to how to quantify operability; instead, the concept "ease of performance" is introduced.

Paper [3] briefly discusses some ways of visualising operability-related, measurable state variables in a wheel loader cycle. Amongst others, a diagram displaying bucket height over integrated machine speed (i.e. the machine's travelling distance) is introduced (Figure 4).

The ratio of lift speed to machine speed, which is crucial to how long the loader needs to be driven backwards until reversing, can here be seen as the curve's slope.

In the diagram in Figure 4 the curve is fairly straight from the beginning of phase 2 (where the loader leaves the bank) to the end of phase 5 (where the loader arrives at the load receiver). This indicates that the operator judged very well when to reverse. Otherwise, the slope would become steeper at the end of phase 5, indicating that the operator needed to slow down or even stop the loader in order for the bucket to reach sufficient height for emptying.

The distance between the point of reversing and the bank (or load receiver) is an indication of how well a balance between the two motions of lifting and driving has been achieved. Because such a diagram points out any imbalance, engineers at Volvo CE have begun to call it *Machine harmony diagram*.

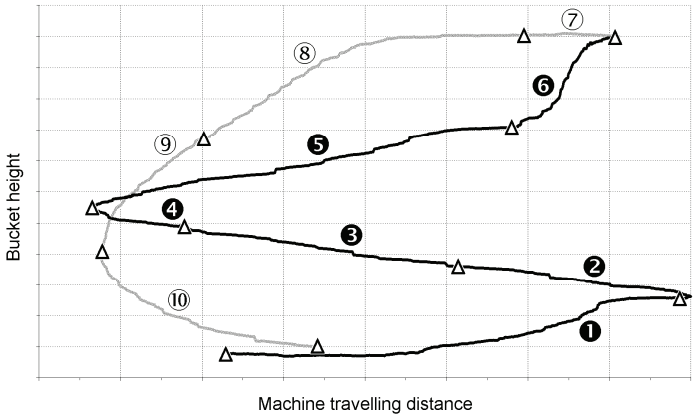

**Figure 4.** Bucket height over integrated machine speed in a short loading cycle

Another interesting aspect is the visualisation of the operator's technique for bucket filling and bucket emptying. For this, the machine harmony diagram is preferably accompanied by another diagram displaying bucket height over bucket angle (Figure 5). This additional diagram also indicates the loader linkage's capability for parallel alignment (i.e. how much the bucket angle changes during raising and lowering of the loading unit).

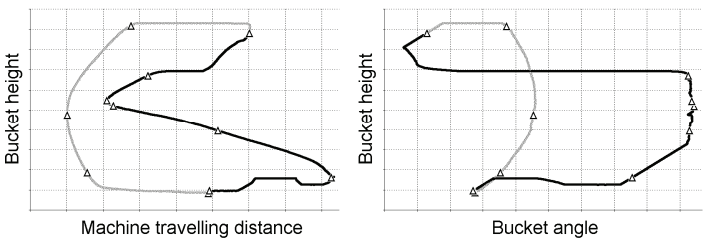

**Figure 5.** Example of bucket handling in a short loading cycle

The chosen bucket filling technique indirectly influences where the loader needs to reverse: Leaving the bank at the beginning of phase 2 with a higher bucket means that it will take less time to raise it from there to a sufficient height for emptying. As explained before, an experienced operator chooses the point of reversing so that the bucket will have the necessary height approximately when the loader reaches the load receiver. This means that in theory the point of reversing will be nearer to the bank when leaving with a higher bucket position, because the operator adapts to the machine.

It can be easily shown that this is the case also in practice. Figure 6 shows three selected short loading cycles, originating from a non-stop recording of a longer period, during which the operator changed his bucket filling technique. As can be seen, the bucket height when leaving the bank varies, but the operator managed very well to chose a reversing point in order to reach the load receiver with the bucket at a sufficient height for emptying.

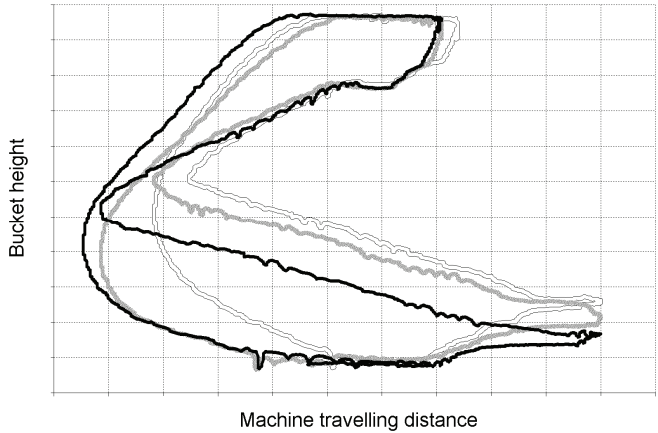

**Figure 6.** Recorded short loading cycles with different bucket filling techniques

Again, it is this kind of human adaptation to the machine which is the reason for our work on operator models: we want to be able to assess a new machine's productivity, efficiency, operability, and their robustness by means of simulation.

## *3*  Simulation Setup

In previous work, the author and two colleagues modelled a wheel loader in a simulation suite for continuous multi-body simulation. The model features a detailed hydraulic system, connected to a fairly detailed mechanical model (rigid bodies) and drive train. The engine, with engine controller and gas dynamics, was realised as state variables, enhanced by user-written subroutines that are linked to the numerical solver. The environment is represented as a model of a gravel pile [5], and as a standard tyre/road model.

Since using the loader model for the simulations presented in [2], the interface to the operator model has been enhanced. Already in the previous version, the operator model was logically separated from the machine model, which was controlled through variables representing the functions engine throttle, brake, steering, lift, and tilt. However, everything was still located within one model space. Now, by using co-simulation, the separation is complete: the wheel loader is modelled and simulated in the multi-body simulation program, while the operator is completely modelled and simulated in the previously mentioned package that supports both continuous and discrete event simulation. The operator model sends commands through the machine controls, which the loader model receives, uses and answers with e.g. loader position and orientation, and engine speed (Figure 7). Some additional elements have been added to simplify debugging.

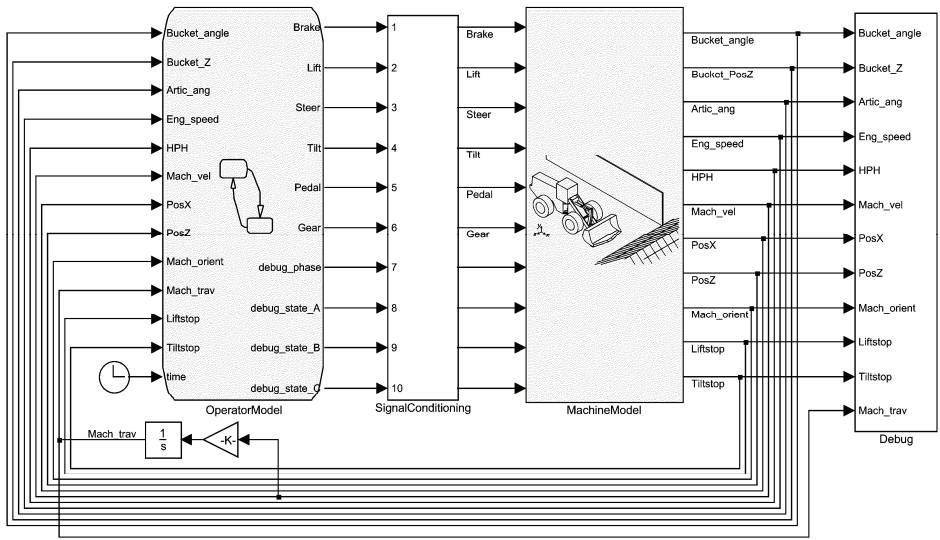

**Figure 7.** Simulation setup (overall view)

Both sides act as black box models towards each other. No internal states are sent, because the intention is to model human-machine interaction, which naturally occurs through a limited number of channels. As long as the interface is unchanged, the models on either side can be replaced by just copying a different file into the simulation's work directory.

In the near future, our ambition is to compile operator models into executable files that can be imported as General State Equations into the multi-body simulation program. This will significantly simplify the simulation process. Initial tests in that respect have shown very promising results.

## 4 Event–based Operator Model

Figure 8 shows the overall view of the operator model, where each phase has been modelled as a state containing several substates.

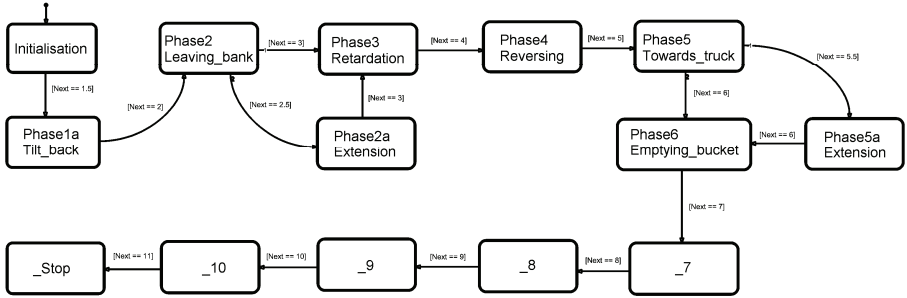

**Figure 8.** Operator model (top level view)

Paper [2] mainly focussed on modelling one bucket filling technique, with encouraging results. Therefore, the operator model presented in this paper has been designed to begin where the bucket filling phase in [2] ended, immediately before tilting back the bucket and leaving the gravel pile. An initialisation phase has been modelled (Figure 9), during which the bucket is lifted and tilted to the same position that resulted in [2]. The engine is also set to the same speed. Already during model preparation the loader has been moved to the same position it had at the end of bucket filling in [2].

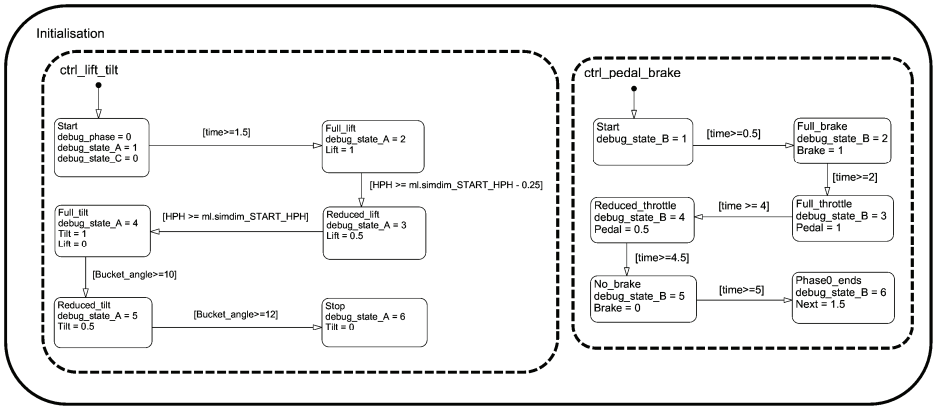

**Figure 9.** Initialisation phase

As described earlier, at the end of phase 1, *Bucket filling*, the operator tilts the loader's bucket back as far as possible. In the model, this is done in the intermediate phase 1a.

In the next phases the operator will begin by calculating and following a theoretical path, which is only defined by the location of bank and load receiver. Later on, the operator model will diverge from this path to react to certain events. For all calculations of position and orientation, a global co-ordinate system with its origin in the intersection of bank and load receiver is used (Figure 10a).

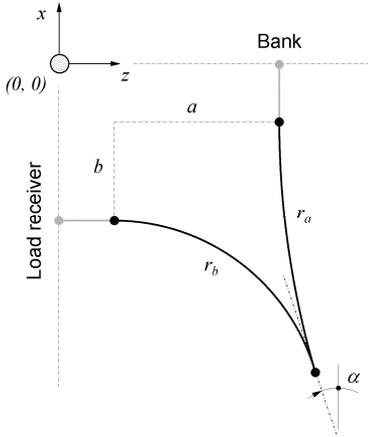
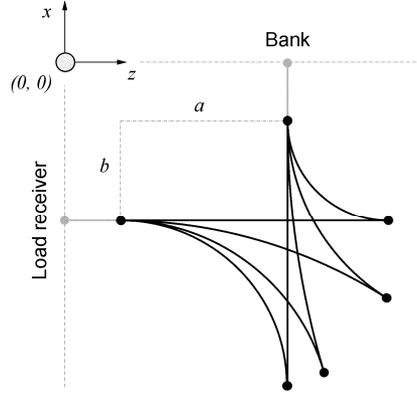

**Figure 10a.** V-pattern, theoretical       **Figure 10b.** Theoretical path solutions

The theoretical path begins with a straight line that is perpendicular to the bank. It also ends with a straight line, this one perpendicular to the load receiver. In between, the path consists of two circular arcs that are tangential to each other and the two lines. Such a geometrical figure has one degree of freedom, which can be expressed as the angle of the tangent that is shared by the arcs. Figure 10b shows four possible solutions.

Which solution to take is dependent on how much space the workplace offers. For a given angle $\alpha$ the circular arcs' radii $r_a$ and $r_b$ result in:

$$r_a = \frac{1}{2}\left(\frac{(a+b)(1+\cos\alpha)}{\sin\alpha} - (a-b)\right) \tag{1}$$

$$r_b = \frac{1}{2}\left(\frac{(a+b)\cos\alpha}{(1-\sin\alpha)} + (a-b)\right) \tag{2}$$

Many operators prefer to steer the loader approximately equally much to the right as to the left. This corresponds to Figure 3 in which the global tangent orientation is stated to be $\alpha \approx 45°$. Equations 1 and 2 can then be simplified to:

$$r_a = \frac{a+b}{\sqrt{2}} + b \tag{3}$$

$$r_b = \frac{a - b\left(1 - \sqrt{2}\right)}{2 - \sqrt{2}} \tag{4}$$

However, for the model presented in this paper, the reasoning was that an operator prefers to aim at a fixed point while reversing. This is easier to accomplish and also results in a 45 degree angle when the global origin is used as fixed point and when, as in most cases, load receiver and bank are located equidistantly from the global origin. The circular arcs' radii and the global tangent orientation can be calculated as:

$$r_a = b + \frac{(a+b)\sqrt{(a+b)^2 + 4ab} - (a^2 - b^2)}{4a} \tag{5}$$

$$r_b = a + \frac{(a+b)\sqrt{(a+b)^2 + 4ab} + (a^2 - b^2)}{4b} \tag{6}$$

$$\cos\alpha = \frac{a + r_a}{r_a + r_b} \tag{7}$$

The case of "aiming at the origin" can easily be checked by comparing $\beta$, the loader's bearing to the origin with $\theta$, its orientation in the global co-ordinate system (Figure 11).

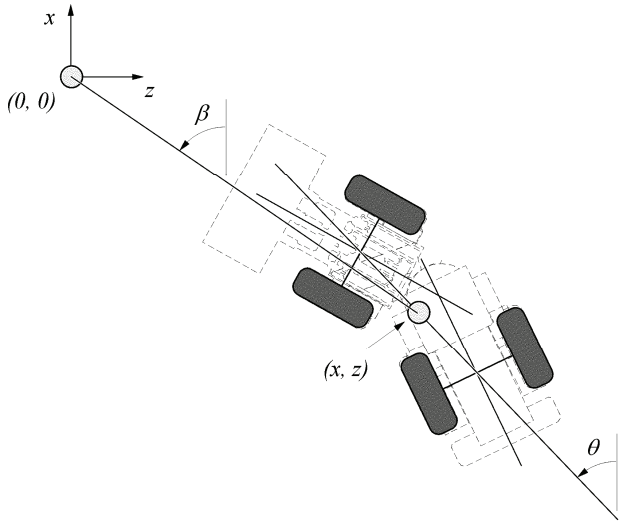

**Figure 11.** Global position, orientation $\theta$ and bearing $\beta$

Phase 2, *Leaving bank,* begins with putting the transmission into reverse. Then, full lift and the machine's steering function are activated until an articulation angle is reached that corresponds to a radius according to Eq. 5 (i.e. plan to reverse with aim at origin). However, due to the lack of feed-back path control in this version of the operator model, the machine model will not follow the theoretical path exactly. But this is not necessary either, because another rule is activated when the machine prematurely aims at the origin. This rule overrides the original path and steers the machine to drive along the new-found bearing line.

As soon as the loader passes the load receiver, an estimation routine begins to continuously extrapolate the current lifting/driving ratio in order to determine whether the reversing phase can be entered. The remaining distance to the load receiver is calculated as a combination of a circular arc with radius $r_c$ and a straight line (Figure 12).

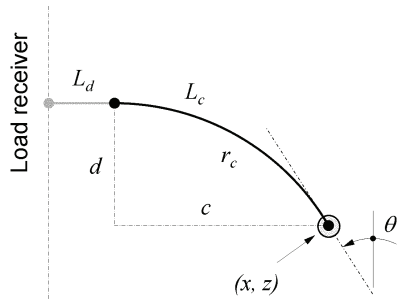

**Figure 12.** Path to load receiver

Knowing the machine's current position $(x, z)$ and global orientation $\theta$, the turning radius $r_c$ and the remaining travelling distance $L$ can be found as:

$$r_c = \frac{d}{1-\sin\theta} \tag{8}$$

$$L_c = r_c\left(\frac{\pi}{2} - \theta\right) \tag{9}$$

$$L_d = z - r_c \cos\theta \tag{10}$$

$$L = L_c + L_d \tag{11}$$

However, most wheel loaders have a maximum articulation angle of 35-40 degrees and might therefore not be able to achieve the currently required turning radius $r_c$. To handle this situation in a simulation, an intermediate phase 2a is activated. During this phase, the operator model continues the V-pattern from phase 2 and conducts calculations according to Eq. 8-11. As soon as the required turning radius $r_c$ is actually achievable, phase 2a is left. It is advantageous to handle this situation in a special phase; otherwise, an extended phase 2 might be interpreted as a limited lifting speed of the loader rather than a limited turning capability.

Immediately after activation of phase 3, *Retardation*, the engine throttle is released and the articulation steered back to zero. In addition to the engine, the service brakes are also used to decrease the machine's speed.

Phase 4, *Reversing*, begins when the machine's speed has decreased to a level that where setting the transmission to forward is safe. The brakes are released, the engine throttle is activated again, and lifting is stopped until sufficient speed has been built up.

In phase 5, *Towards load receiver*, the required turning radius $r_c$ is again calculated with Eq. 8. However, a theoretically correct but impractical solution to Eq. 8-11 might be found that yields a negative value for $L_d$, the length of the straight line between load receiver and circular arc (Figure 12). One possible strategy is to just move forward in a straight line until the current machine position and global orientation result in a positive value for $L_d$. When the machine finally reaches the load receiver, the articulation angle is adjusted so that the bucket is parallel to the load receiver, regardless of the wheel loaders orientation. However, in a loading cycle that is not too short, the complete wheel loader should arrive with zero articulation and perpendicular to the load receiver.

During phase 5, the lifting function is activated until the bucket is at a sufficient height for emptying. If this is not the case when the machine reaches the load receiver, an intermediate phase 5a is started before the bucket is emptied in phase 6. In this intermediate phase, the operator model is using the lift function to reach a bucket height that is sufficient for emptying into the load receiver.

In phase 6, *Bucket emptying*, the operator model uses engine throttle, lift, and tilt function simultaneously in order to simulate a human operator's strategy of placing material so that the load receiver is evenly loaded after three to four short loading cycles.

As explained earlier, a short loading cycle continues with phases 7-10, *Leaving load receiver*, *Retardation and reversing*, *Towards bank*, and *Retardation at bank*, which are not covered yet in the presented operator model.

## 5   Simulation Results

In order to give answers with respect to a simulated wheel loader's productivity, efficiency, and operability, the operator model needs to be updated to include all phases of a short loading cycle, most importantly bucket filling. Until this is done, simulation results in respect of engine load duty and fuel consumption should be considered with care. For this reason this paper will only present how the operator model succeeded in adapting to variations in workplace layout and machine properties.

Figure 13 shows the results of a short loading cycle that has the same workplace layout as the cycle presented in paper [2]. Compared to the model in [2], this one represents an operator with a very rough operating style; most control inputs are either on or off, and the engine throttle is very often at maximum value.

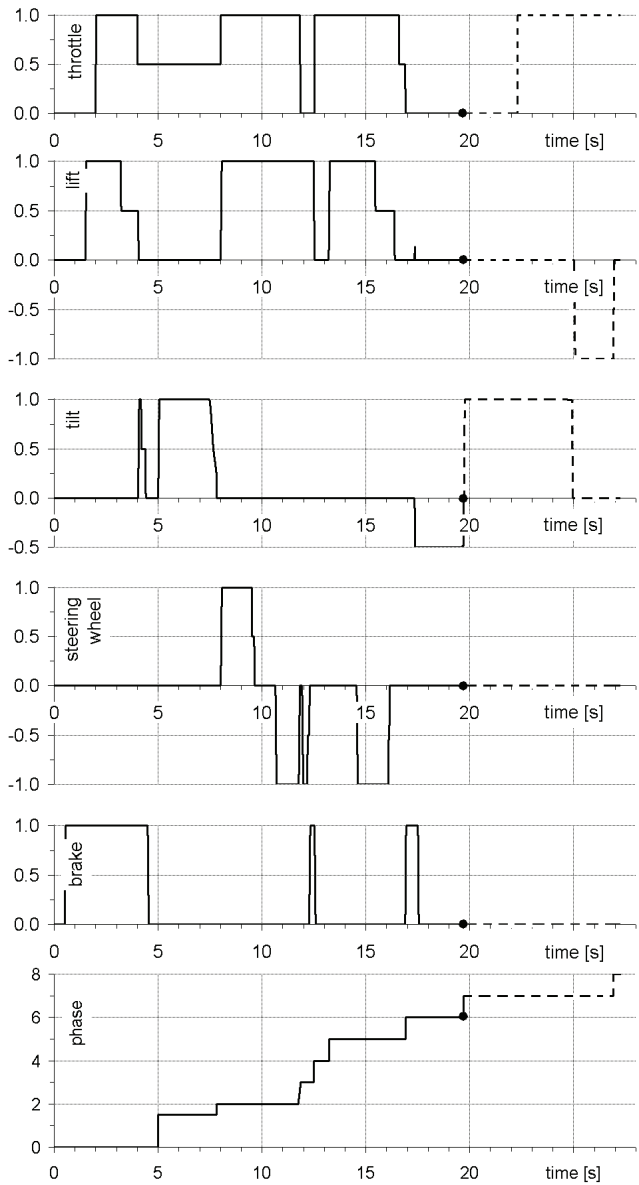

**Figure 13.** Control input signals from the operator model

The diagram on the bottom displays the current cycle phase (#0 is the initialisation phase).

Figure 14 displays the machine movement from the cycle above in both a location plot and a harmony diagram. Since phase 1, *Bucket filling*, is not included in the current version of the operator model, the cycle in the harmony diagram starts with a vertical line that shows how the loader model lifts the bucket in the simulation's initialisation phase.

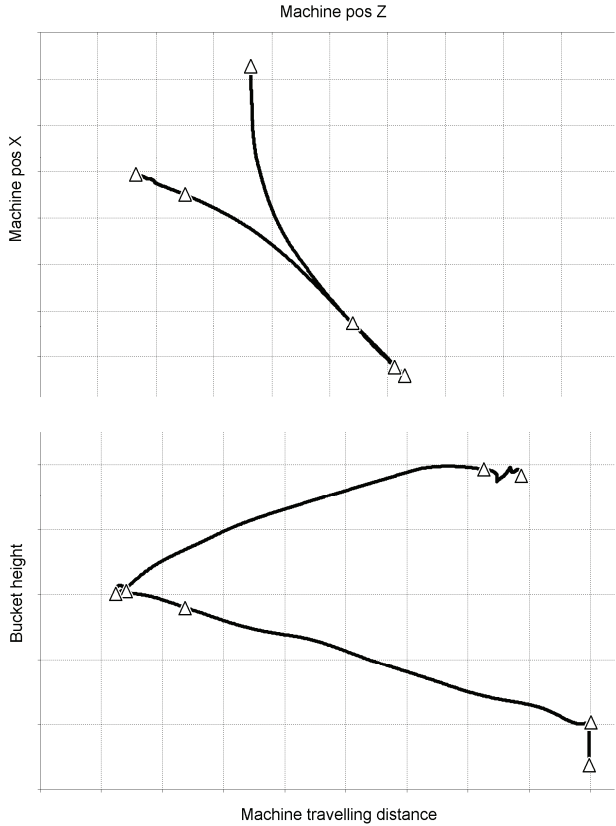

**Figure 14.** Machine location plot and harmony diagram

Because the operator model has not been hard-coded with fixed locations for bank and load receiver, but designed parametrically, it is able to adapt. Figure 15 shows this for two different load receiver positions.

In the initial simulation setup (Figure 13 and Figure 14) the machine's movement was more restricted by its limited turning capability, rather by its lifting speed. The simulated wheel loader's control signal has therefore been artificially decreased by 50%, which results in a slower lifting speed. Figure 16 shows how the operator model adapts to this situation by choosing a different reversing point.

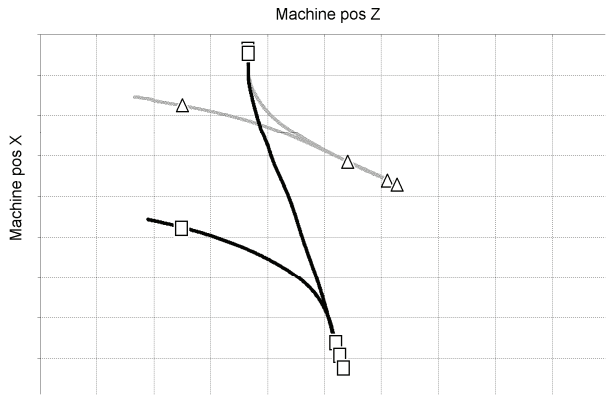

**Figure 15.** Adaptation to workplace layout

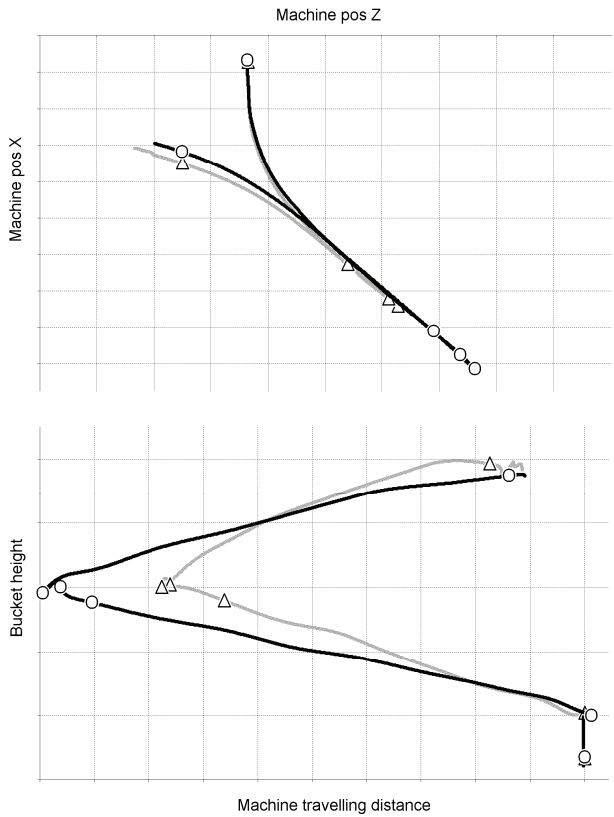

**Figure 16.** Adaptation to lifting speed

Finally, as already explained earlier and shown in Figure 6, an operator's bucket filling technique has an indirect influence on the location of the reversing point. With a higher bucket position at the beginning of phase 2, the point of reversing can be chosen nearer to the load receiver. Figure 17 presents results of two simulations that start with different bucket heights (in both cases the loader model with artificially decreased lifting speed was used).

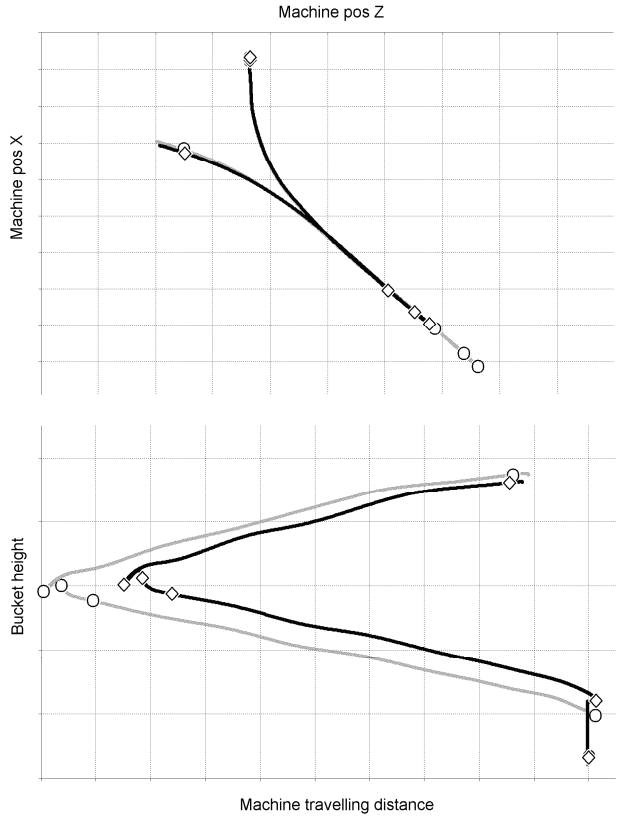

**Figure 17.** Influence of bucket filling technique

## 6   Discussion

While the operator model presented in paper [2] focussed on the bucket filling phase and used min/max relationships reminiscent of fuzzy-sets, the operator model presented here focuses on the remaining elements of a short loading cycle and uses a pure discrete-event approach. It is advantageous to use this modelling paradigm, but certain elements such as path control are probably best modelled with a simple PI-controller. Therefore, in the next version of this operator model the possibility of combining these approaches needs to be explored. Because the chosen simulation package supports both continuous and discrete-event simulation, this is a feasible prospect.

As mentioned earlier, one interesting aspect is the possibility of compiling the operator model into an executable file, which is referenced as a General State Equation in the multi-body simulation program. With this, the engineer conducting a simulation can focus on the wheel loader, rather than having to spend time defining how it needs to be controlled to mimic reality.

# 7   Conclusion

An event-driven operator model has been developed, which is able to adapt to basic variations in workplace layout and machine capability. With this, a "human element" can be introduced into dynamic simulation of complete wheel loaders, giving more relevant answers with respect to total machine performance, fuel efficiency, and possibly even operability in complete loading cycles.

## Acknowledgements

The financial support of Volvo Wheel Loaders AB and PFF, the Swedish Program Board for Automotive Research, is hereby gratefully acknowledged.

(Internet links updated and verified on August 18, 2011)